\documentclass[aps,prl,twocolumn]{revtex4-2}

\usepackage{amsmath,amssymb}
\usepackage{graphics}

\newcommand{\eqtri}{\triangleq}
\newcommand{\avg}[1]{\langle{#1}\rangle}
\newcommand{\Dcirc}{\mathcal{D}^{\circ}}

\begin{document}

\title{Nature abhors macroscopic superpositions}

\author{Filippus S. Roux}
\email{rouxf@ukzn.ac.za}
\affiliation{University of KwaZulu-Natal, Private Bag X54001, Durban 4000, South Africa}

\begin{abstract}
Superpositions of mass distributions can potentially lead to entanglement with the geometry of spacetime. Here we show that there exists a natural reluctance for macroscopic mass distributions to form such superpositions. The macroscopic superposition is modelled as a Schr{\"o}dinger cat state. The reluctance manifests as a dip in the total energy of the Schr{\"o}dinger cat state as a function of the separation distance between the terms in the superposition. The dip in the energy provides an opposing force preventing the formation of the superposition. A generalization of this phenomenon addressing the measurement problem is also discussed.
\end{abstract}

\maketitle

\begin{figure}[ht]
\centerline{\includegraphics{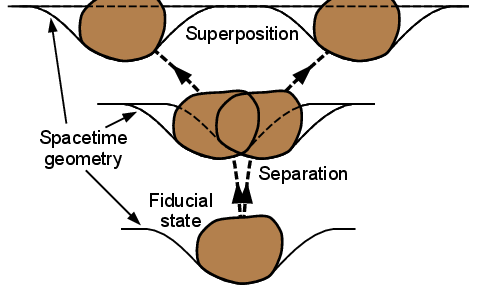}}
\caption{The fiducial state of a mass distribution evolves by separating into a superposition of mass distributions located at two different positions. }
\label{skeifig}
\end{figure}

Among physical scenarios involving both gravity and quantum physics, we are confronted with the question of how a superposition of massive objects, as shown in Fig.~\ref{skeifig}, would affect the geometry of spacetime. Would such a superposition cause entanglement between its terms and the geometry of spacetime if these terms are heavy enough and spread far enough apart to affect the geometry differently via gravity? We never see such situations. Why not? The answer to these questions may significantly contribute to a fundamental understanding of the coexistence of gravity and quantum physics.

A different question that is somehow related to the previous questions is the so-called \emph{measurement problem} \cite{schlosshauer}. According to our current understanding, a measurement device with two possible measurement outcomes evolves into a superposition of the two states associated with these outcomes \cite{wm}. How does such a superposition lead to the single measurement outcome that we expect?

To address these questions, some people presented proposals calling for the introduction of novel principles in physics that would prevent such superpositions. For example, it was proposed that there may be an instability mechanism \cite{karolyhazy,diosi1,diosi2,penrose0,penrose} in which some of the terms of the superposition shown in Fig.~\ref{skeifig} would decay at a time scale governed by the amount of action given by Planck's constant. This idea was proposed in part to address the measurement problem.

Here, we first focus on the issue of a superposition of mass distributions that can potentially become entangled with the geometry of spacetime. We also consider a possible mechanism that can prevent such superpositions from becoming entangled with the spacetime geometry, but instead of an instability mechanism (which would only kick in once the superposition has been produced), we consider a variational principle that would prevent superpositions from being formed in the first place. Such a variational principle can limit the spread of superpositions in a quantum state. The term ``spread'' represents a measure of the separation among the different terms in such a superposition of mass distributions. In the context of a possible entanglement with spacetime geometry, the spread represents the distances among the physical spacetime locations of the terms in the superposition of mass distributions.

A variational principle is enforced by a quantity (an observable) that produces a varying landscape over the parameter space of the quantum state associated with a given object. The idea is that minima in the landscape of such a quantity would introduce a mechanism that limits the spread of a superposition.

How would such a variational principle be incorporated into existing theories? One may expect that the quantity associated with such a variational principle would have to be included as an additional term in the theory if it is not already included. So, if a new term is required to implement the variational principle, it may lead to modifications of the theory's predictions that may contradict previous observations. If such a variational principle does exist, there must be a reason why we have not seen such deviations yet. This issue would not occur if the quantity that is associated with such a variational principle already exists in the theory. As it turns out, the required variational mechanism is indeed manifested by the total energy of the superposition, which does not require any modification to the existing dynamics of any theory. The calculation of the energy in a quantum state is not too complicated. The only complication that we need to introduce in our calculations is the use of a functional approach that allows the required freedom in the parameterization of quantum states.

To investigate such a variational principle, one also needs to provide a model for the superposition on which this variational principle should take effect. For the possible entanglement with spacetime, we expect to see this mechanism for a superposition of \emph{translated} mass distributions. To model it, we consider a Schr{\"o}dinger cat state \cite{nc} where the terms represent oppositely translated mass distributions, each represented by the thermal state of a massive scalar field.

Our calculation demonstrates that the total energy of a Schr{\"o}dinger cat state has a minimum as a function of the separation distance close to the origin at zero separation. The minimum is located at smaller separation distances for larger numbers of particles in the state. The energy curve serves as a potential function producing a force that opposes any increase in the separation distance.

Although this scenario is very simple, it has a profound implication. Macroscopic states have large numbers of particles leading to dips so close to the origin that their locations cannot be distinguished from those without superpositions. Moreover, the force opposing the formation of a superposition is so large that it becomes virtually impossible to produce superpositions with discernable spreads. This observation does not only explain why entanglement between mass distributions and the geometry of spacetime is not observed, but also provides a possible explanation why macroscopic measurements collapse to produce single measurement outcomes. We return to this possibility after we have discussed the calculation of the energy curves.


While a Schr{\"o}dinger cat state is the obvious choice for modeling a superposition of mass distributions, there are different ways to distinguish the different terms in the superposition. Often these terms are distinguished by different \emph{displacements} as produced by displacement operators. However, displacement operators nominally affect the particle number degree of freedom. To obtain a superposition of mass distributions that can potentially become entangled with spacetime, we need to distinguish the terms by \emph{translations}, which affect the spatial degrees of freedom. Such translations are accomplished by phase modulations of the parameter function of the state, requiring a phase operator. If a state is parameterized in the Fourier domain by a spectral function $\zeta(\mathbf{k})$ where $\mathbf{k}$ is the wave vector, a translation is readily implemented by multiplying it with a linear phase factor
\begin{equation}
\zeta(\mathbf{k}) \rightarrow \exp(i\mathbf{x}_0\cdot\mathbf{k}) \zeta(\mathbf{k}) .
\end{equation}
Here $\mathbf{x}_0$ is the spatial translation vector. In what follows, we assume translation along the $z$-direction.

We also need to decide what kind of state to employ as the fiducial state on which the translations are applied. A coherent state is often used in the definitions of Schr{\"o}dinger cat states. However, any kind of state may serve as the fiducial state, depending on the nature of the physical scenario. Here, we consider a single-mode thermal state. It represents a mixed state with a parameter function describing the spatial mass distribution. (In more general physical scenarios, the parameter functions and the spatial mass distributions may be different functions having vastly different scales.)

The scenario thus presented is relatively simple. To model more general scenarios, it is preferable to present the basic calculation within a formalism that readily lends itself to such generalizations. Therefore, we perform the calculation on a functional phase space \cite{mrowc,fpsm}, allowing representations of states and operations in all degrees of freedom. The measurement of the energy is given by a trace of the product of the Wigner functionals of the state and the energy observable. In the context of a free theory, the latter reads
\begin{equation}
W_{\mathcal{E}} = \alpha^*\diamond\mathcal{E}\diamond\alpha-\tfrac{1}{2} \Omega_{\mathcal{E}} ,
\end{equation}
where $\alpha(\mathbf{k})$ is the complex field variable of the functional phase space, the $\diamond$-contraction is a short-hand for an integral over a shared wave vector $\mathbf{k}$, the kernel is
\begin{equation}
\mathcal{E}(\mathbf{k},\mathbf{k}') \eqtri \hbar\omega_{\mathbf{k}} \mathbf{1}(\mathbf{k},\mathbf{k}') ,
\end{equation}
where
\begin{equation}
\mathbf{1}(\mathbf{k},\mathbf{k}') \eqtri (2\pi)^3 \delta(\mathbf{k}-\mathbf{k}') ,
\end{equation}
and $\Omega_{\mathcal{E}}$ is a divergent constant (the zero-point energy). It serves to remove a similar divergent constant from the calculated energy.

Functional integrals of the Wigner functional of a state multiplied by powers of the field variable produce the different moments of the Wigner functional. Such moments are readily obtained from the characteristic functional of the state (the symplectic functional Fourier transform of the Wigner functional). The characteristic functional is a generating functional for moments of a state's Wigner functional. Therefore, we can represent the measurement of the energy of a state directly in terms of its characteristic functional $\chi[\eta]$ by
\begin{equation}
\avg{\mathcal{E}} = \left. -\left(\delta_{\eta}\diamond\mathcal{E}\diamond\delta_{\eta}^*
+\tfrac{1}{2} \Omega_{\mathcal{E}} \right)\chi[\eta] \right|_{\eta=0} ,
\label{ekontr}
\end{equation}
where $\delta_{\eta}$ and $\delta_{\eta}^*$ represent functional derivatives with respect to the auxiliary field variable of the characteristic functional $\eta$ and its complex conjugate, respectively.

The thermal state's characteristic functional reads
\begin{equation}
\chi_{\text{th}}[\eta] = \exp\left(-\tfrac{1}{2}\eta^*\diamond\theta^{-1}\diamond\eta\right) ,
\label{karterm}
\end{equation}
where $\theta(\mathbf{k},\mathbf{k}')$ is the thermal state kernel. Here we assume that the fiducial state is a single-mode thermal state. Its inverse kernel is given by
\begin{equation}
\theta^{-1}(\mathbf{k},\mathbf{k}') = \mathbf{1}(\mathbf{k},\mathbf{k}')
+2\avg{n}\Theta(\mathbf{k}) \Theta^*(\mathbf{k}') ,
\end{equation}
where $\avg{n}$ is the average number of particles in the state and $\Theta(\mathbf{k})$ is the thermal state mode, modeling the mass distribution in terms of its Fourier spectrum.

For the characteristic functional of the Schr{\"o}dinger cat state, we combine the fiducial state's characteristic functional with that of the translation operator, using the star product of characteristic functionals. The characteristic functional of the translation operator is given by
\begin{equation}
\chi_{\hat{U}}[\eta] = \mathcal{N}_{\hat{U}} \exp(-\eta^*\diamond\tau\diamond\eta) ,
\end{equation}
where $\mathcal{N}_{\hat{U}}=1/\det\{\mathbf{1}-\Phi\}$ and
$\tau=\tfrac{1}{2}(\mathbf{1}+\Phi)\diamond(\mathbf{1}-\Phi)^{-1}$, with $\Phi=\exp(i z k_z)\mathbf{1}$, and
the functional integral for the star product of characteristic functionals reads
\begin{align}
\chi_{\hat{A}} \star \chi_{\hat{B}} = & \int \chi_{\hat{A}}[\beta] \chi_{\hat{B}}[\eta-\beta] \nonumber \\
&\times \exp(\tfrac{1}{2}\eta^*\diamond\beta-\tfrac{1}{2}\beta^*\diamond\eta)\ \Dcirc[\beta] ,
\label{sterkar}
\end{align}
where $\Dcirc[\beta]$ is the functional integration measure. The resulting characteristic functional of
the Schr{\"o}dinger cat state is
\begin{align}
\chi_{\text{cat}}[\eta] = &\ \frac{1}{2(1+\Lambda)} \exp\left(-\tfrac{1}{2}\eta^*\diamond\eta\right) \nonumber \\
&\times \left[
\exp\left(-\avg{n}\eta^*\diamond\Phi\diamond\Theta\Theta^*\diamond\Phi^*\diamond\eta\right)\right. \nonumber \\
& +\Lambda\exp\left(-\avg{n}\Lambda\eta^*\diamond\Phi\diamond\Theta\Theta^*\diamond\Phi\diamond\eta\right) \nonumber \\
& +\Lambda\exp\left(-\avg{n}\Lambda\eta^*\diamond\Phi^*\diamond\Theta\Theta^*\diamond\Phi^*\diamond\eta\right) \nonumber \\
& \left. +\exp\left(-\avg{n}\eta^*\diamond\Phi^*\diamond\Theta \Theta^*\diamond\Phi\diamond\eta\right)\right] ,
\label{kartraterm}
\end{align}
where, we define
\begin{equation}
\Lambda \eqtri \frac{1}{1+\avg{n}-\mu} ,
\end{equation}
assuming that $\Theta$ is symmetric, so that
\begin{equation}
\mu\eqtri\Theta^*\diamond\Phi^2\diamond\Theta=\Theta^*\diamond\Phi^{*2}\diamond\Theta .
\end{equation}


Applying the process given in Eq.~(\ref{ekontr}) to Eq.~(\ref{kartraterm}), we obtain the energy in the Schr{\"o}dinger cat state
\begin{align}
\avg{\mathcal{E}}_{\text{cat}}
= &\ \frac{\avg{n}}{2(1+\Lambda)}\left(2\Theta^*\diamond\mathcal{E}\diamond\Theta
+\Lambda^2 \Theta^*\diamond\Phi^2\diamond\mathcal{E}\diamond\Theta \right. \nonumber \\
&\ \left. +\Lambda^2 \Theta^*\diamond\Phi^{*2}\diamond\mathcal{E}\diamond\Theta \right) .
\end{align}
Since both $\mathcal{E}$ and $\Phi$ are diagonal kernels, they commute. For the calculation of the $\diamond$-contractions, we need to define a function for the thermal state mode that models the mass distribution. However, since the dominant physical effect is provided by the scale (or width) of the mass distribution, we simply assume a normalized Gaussian function. It leads to
\begin{equation}
\mu = \avg{n}\exp\left(-\frac{4z^2}{w_0^2}\right) \eqtri \avg{n} g(z) ,
\label{defgz}
\end{equation}
where $w_0$ is the parameter representing the width of the mass distribution. Moreover
\begin{equation}
\Theta^*\diamond\Phi^2\diamond\mathcal{E}\diamond\Theta
= \Theta^*\diamond\Phi^{*2}\diamond\mathcal{E}\diamond\Theta
= \mathcal{E}_0^{(\text{th})} g(z) ,
\end{equation}
where
\begin{equation}
\mathcal{E}_0^{(\text{th})} \eqtri \Theta^*\diamond\mathcal{E}\diamond\Theta.
\end{equation}
The energy of the fiducial state is $\avg{n} \mathcal{E}_0^{(\text{th})}$.

The energy in the Schr{\"o}dinger cat state, as a function of the translation distance $z$, thus becomes
\begin{equation}
\avg{\mathcal{E}}_{\text{cat}} = \avg{n} \mathcal{E}_0^{(\text{th})}
\left[ 1-\frac{(\avg{n}+1)h(z)}{2+3\avg{n}h(z)+\avg{n}^2 h^2(z)} \right] ,
\end{equation}
where $h(z)=1-g(z)$.

\begin{figure}[ht]
\centerline{\includegraphics{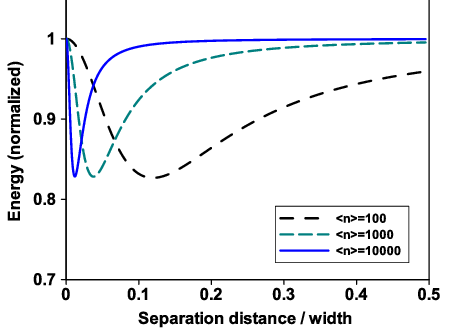}}
\caption{The normalized energy of the Schr{\"o}dinger cat state is plotted for $\avg{n}=10^2, 10^3, 10^4$ as a function of the normalized separation distance between the two mass distributions in the superposition.}
\label{dipfig}
\end{figure}

In Fig.~\ref{dipfig}, we plot curves of the normalized energy of the Schr{\"o}dinger cat state as a function of the normalized separation distance $2z/w_0$ for different values of the average number of particles $\avg{n}$. The curves have dips next to the origin. The dips become narrower and move closer to the origin with larger numbers of particles, but their relative depths remain more or less the same. The opposing force therefore increases with increasing numbers of particles.

Any macroscopic physical object is expected to find itself at the bottom of the dip, having progressed toward it adiabatically during whatever process that formed the object. With the dip located next to the origin the minimum energy state at the bottom of the dip is already a superposition. However, for macroscopic physical objects, this superposition represents such a small difference between the terms that it would be challenging to observe and unlikely to cause any physical effect that would differ from any classical observations. Being located at the bottom of the dip, such a macroscopic object opposes any attempt to turn it into a larger more visible superposition. For small numbers of particles, such superpositions are allowed to consist of terms that are more distinct, which may allow experimental observations of the dip.

While we use the single-mode thermal state to provide a simple tractable model for a superposition of mass distributions, it is not the only state that produces a dip close to origin when used in a superposition based on translations. A coherent state also produces such a dip but the relative depth decreases with increasing numbers of particles. Nevertheless when the normalized energy curves are multiplied by the absolute energy at the origin, the depths of these dips also increase but at a slower rate compared to those of the thermal states.

Here we formed the Schr{\"o}dinger cat state by applying translations to the fiducial state. In scenarios that are relevant for superpositions produced in measurements, the unitary transformations are better modelled as displacements. The energy curves of Schr{\"o}dinger cat states that are produced with displacements of coherent states and thermal states also have dips close to (or at) the origin. The parameter space becomes larger when considering different possible scenarios, but all those cases that we investigated produce dips in their energy curves as a function of a free parameter that smoothly varies from the fiducial state to the Schr{\"o}dinger cat state. These curves represent potential functions, but since the continuous parameter is not a distance, the derivative is not a force. Yet, it still provides a mechanism opposing any process that attempts to evolve the fiducial state into the Schr{\"o}dinger cat state.

\begin{figure}[ht]
\centerline{\includegraphics{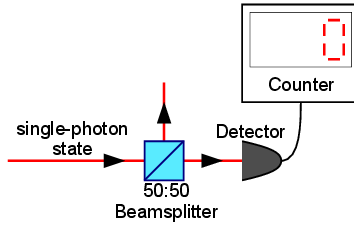}}
\caption{Measurement setup for single-photon detection after a beamsplitter. }
\label{meetfig}
\end{figure}

The theory of quantum measurements \cite{wm} calls for a process where the tensor product of the state of a measurement device and the state to be measured evolves into a superposition of pointer states representing the different outcomes of the measurement. For example, consider a photon detector coupled to a counter as shown in Fig.~\ref{meetfig}. The photon detector receives either zero or one photon, leading to two possible measurement outcomes, respectively producing the numbers ``0'' or ``1'' on the display of the counter. Unless some process of quantum collapse occurs, the measurement device would be in a superposition of these two measurement outcomes. However, assuming that the evolution process of the measurement is governed by some continuous parameter $\lambda$, the above calculation suggests that we'll find a dip in the total energy of the system as a function of $\lambda$. Instead of a superposition of the two measurement outcomes, we have a situation where the system either evolves deterministically toward ``0'' or toward ``1'' with the opposite term being ``dragged along'' due to it being located at the bottom of the dip. The dominant outcome is determined by the magnitude of the probability amplitudes for the two outcomes. (In a continuous probability distribution, the probability for these two magnitudes to be equal is zero.)

In the calculation above, we used the energy operator as defined by the normal-ordered free Hamiltonian to serve as the variational quantity. However, the variational quantity can be any of a variety of observables. (Since the kernel of the energy operator does not play a crucial role in the formation of the dip, it can be replaced by any other kernel without changing the qualitative result.) In the presence of interactions, there would be interaction terms in addition to the free Hamitonian term. For example, one can for example consider a four-point interaction or any generalization of it. The effect of such a quantity (applied to the state considered above) was also found to produce complicated oscillating curves with multiple dips. Those close to (or at) the origin follow the same behaviour as found for the energy. The details of the four-point kernels do not play a significant role in the qualitative properties of these dips.

The shapes of these dips and the way they change with increasing numbers of particles are different for the different scenarios. However, they all have the same generic properties: (a) The dips move closer to the origin as the number of particles increases and may even be located at the origin at some point. (b) The slope of the opposing side of the dip in the variational quantity's curve, which represents the opposing force in the case of a separation distance, always increases for increasing number of particles in the fiducial state.

This investigation, although far from being exhaustive, gives a strong indication that the appearance of dips in the curves of the energy (or similar variational quantities) of macroscopic superpositions is a generic feature. The generic nature of this feature provides strong support for the conclusion that macroscopic objects will always be reluctant to form superpositions that would convert them into Schr{\"o}dinger cat states.

As much as this investigation provides a convincing indication for the formation of the dips, it would only provide a scientific understanding if the appearance of such dips can be confirmed in experimental observations. Fortunately, there have recently been several experimental demonstrations of Schr{\"o}dinger cat states
\cite{wineland,grangier,schoelkopf,lvovsky,lewenstein}. While not all of these experimental implementations may be suitable for the observation of such dips, there are scenarios where these dips may be more readily observed. It is hoped that this work will inspire these groups to look for observations that may indicate the appearance of such dips in their systems.


\begin{thebibliography}{15}
\expandafter\ifx\csname natexlab\endcsname\relax\def\natexlab#1{#1}\fi
\expandafter\ifx\csname bibnamefont\endcsname\relax
  \def\bibnamefont#1{#1}\fi
\expandafter\ifx\csname bibfnamefont\endcsname\relax
  \def\bibfnamefont#1{#1}\fi
\expandafter\ifx\csname citenamefont\endcsname\relax
  \def\citenamefont#1{#1}\fi
\expandafter\ifx\csname url\endcsname\relax
  \def\url#1{\texttt{#1}}\fi
\expandafter\ifx\csname urlprefix\endcsname\relax\def\urlprefix{URL }\fi
\providecommand{\bibinfo}[2]{#2}
\providecommand{\eprint}[2][]{\url{#2}}

\bibitem[{\citenamefont{Schlosshauer}(2003)}]{schlosshauer}
\bibinfo{author}{\bibfnamefont{M.}~\bibnamefont{Schlosshauer}}, ``Decoherence,
  the measurement problem, and interpretations of quantum mechanics,''
  \bibinfo{journal}{Rev. Mod. Phys.} \textbf{\bibinfo{volume}{76}},
  \bibinfo{pages}{1267} (\bibinfo{year}{2003}).

\bibitem[{\citenamefont{Walls and Milburn}(1994)}]{wm}
\bibinfo{author}{\bibfnamefont{D.~F.} \bibnamefont{Walls}} \bibnamefont{and}
  \bibinfo{author}{\bibfnamefont{G.~J.} \bibnamefont{Milburn}},
  \emph{\bibinfo{title}{Quantum Optics}} (\bibinfo{publisher}{Springer Verlag},
  \bibinfo{address}{Berlin}, \bibinfo{year}{1994}).

\bibitem[{\citenamefont{Karolyhazy}(1966)}]{karolyhazy}
\bibinfo{author}{\bibfnamefont{F.}~\bibnamefont{Karolyhazy}}, ``Gravitation and
  quantum mechanics of macroscopic objects,'' \bibinfo{journal}{Nuovo Cimento
  A} \textbf{\bibinfo{volume}{42}}, \bibinfo{pages}{390}
  (\bibinfo{year}{1966}).

\bibitem[{\citenamefont{Di{\'o}si}(1987)}]{diosi1}
\bibinfo{author}{\bibfnamefont{L.}~\bibnamefont{Di{\'o}si}}, ``A universal
  master equation for the gravitational violation of quantum mechanics,''
  \bibinfo{journal}{Phys. Lett.} \textbf{\bibinfo{volume}{120A}},
  \bibinfo{pages}{377} (\bibinfo{year}{1987}).

\bibitem[{\citenamefont{Di{\'o}si}(1989)}]{diosi2}
\bibinfo{author}{\bibfnamefont{L.}~\bibnamefont{Di{\'o}si}}, ``Models for
  universal reduction of macroscopic quantum fluctuations,''
  \bibinfo{journal}{Phys. Rev. A} \textbf{\bibinfo{volume}{40}},
  \bibinfo{pages}{1165} (\bibinfo{year}{1989}).

\bibitem[{\citenamefont{Penrose}(1996)}]{penrose0}
\bibinfo{author}{\bibfnamefont{R.}~\bibnamefont{Penrose}}, ``On gravity's role
  in quantum state reduction,'' \bibinfo{journal}{Gen. Rel. Gravit.}
  \textbf{\bibinfo{volume}{28}}, \bibinfo{pages}{581} (\bibinfo{year}{1996}).

\bibitem[{\citenamefont{Penrose}(2014)}]{penrose}
\bibinfo{author}{\bibfnamefont{R.}~\bibnamefont{Penrose}}, ``On the
  gravitization of quantum mechanics 1: Quantum state reduction,''
  \bibinfo{journal}{Found. Phys.} \textbf{\bibinfo{volume}{44}},
  \bibinfo{pages}{557} (\bibinfo{year}{2014}).

\bibitem[{\citenamefont{Nielsen and Chuang}(2000)}]{nc}
\bibinfo{author}{\bibfnamefont{M.~A.} \bibnamefont{Nielsen}} \bibnamefont{and}
  \bibinfo{author}{\bibfnamefont{I.~L.} \bibnamefont{Chuang}},
  \emph{\bibinfo{title}{Quantum Computation and Quantum Information}}
  (\bibinfo{publisher}{Cambridge University Press},
  \bibinfo{address}{Cambridge, England}, \bibinfo{year}{2000}).

\bibitem[{\citenamefont{Mr{\'o}wczy{\'n}ski and Mueller}(1994)}]{mrowc}
\bibinfo{author}{\bibfnamefont{S.}~\bibnamefont{Mr{\'o}wczy{\'n}ski}}
  \bibnamefont{and} \bibinfo{author}{\bibfnamefont{B.}~\bibnamefont{Mueller}},
  ``Wigner functional approach to quantum field dynamics,''
  \bibinfo{journal}{Phys. Rev. D} \textbf{\bibinfo{volume}{50}},
  \bibinfo{pages}{7542} (\bibinfo{year}{1994}).

\bibitem[{\citenamefont{Roux}(2025)}]{fpsm}
\bibinfo{author}{\bibfnamefont{F.~S.} \bibnamefont{Roux}},
  \emph{\bibinfo{title}{Functional Phase Space Methods: Quantum Optics in all
  Degrees of Freedom}} (\bibinfo{publisher}{De Gruyter},
  \bibinfo{address}{Berlin}, \bibinfo{year}{2025}).

\bibitem[{\citenamefont{Monroe et~al.}(1996)\citenamefont{Monroe, Meekhof,
  King, and Wineland}}]{wineland}
\bibinfo{author}{\bibfnamefont{C.}~\bibnamefont{Monroe}},
  \bibinfo{author}{\bibfnamefont{D.~M.} \bibnamefont{Meekhof}},
  \bibinfo{author}{\bibfnamefont{B.~E.} \bibnamefont{King}}, \bibnamefont{and}
  \bibinfo{author}{\bibfnamefont{D.~J.} \bibnamefont{Wineland}}, ``A
  ``{S}chr{\"o}dinger cat'' superposition state of an atom,''
  \bibinfo{journal}{Science} \textbf{\bibinfo{volume}{272}},
  \bibinfo{pages}{1131} (\bibinfo{year}{1996}).

\bibitem[{\citenamefont{Ourjoumtsev et~al.}(2007)\citenamefont{Ourjoumtsev,
  Jeong, Tualle-Brouri, and Grangier}}]{grangier}
\bibinfo{author}{\bibfnamefont{A.}~\bibnamefont{Ourjoumtsev}},
  \bibinfo{author}{\bibfnamefont{H.}~\bibnamefont{Jeong}},
  \bibinfo{author}{\bibfnamefont{R.}~\bibnamefont{Tualle-Brouri}},
  \bibnamefont{and} \bibinfo{author}{\bibfnamefont{P.}~\bibnamefont{Grangier}},
  ``Generation of optical `{S}chr{\"o}dinger cats' from photon number states,''
  \bibinfo{journal}{Nature} \textbf{\bibinfo{volume}{448}},
  \bibinfo{pages}{784–786} (\bibinfo{year}{2007}).

\bibitem[{\citenamefont{Vlastakis et~al.}(2013)\citenamefont{Vlastakis,
  Kirchmair, Leghtas, Nigg, Frunzio, Girvin, Mirrahimi, Devoret, and
  Schoelkopf}}]{schoelkopf}
\bibinfo{author}{\bibfnamefont{B.}~\bibnamefont{Vlastakis}},
  \bibinfo{author}{\bibfnamefont{G.}~\bibnamefont{Kirchmair}},
  \bibinfo{author}{\bibfnamefont{Z.}~\bibnamefont{Leghtas}},
  \bibinfo{author}{\bibfnamefont{S.~E.} \bibnamefont{Nigg}},
  \bibinfo{author}{\bibfnamefont{L.}~\bibnamefont{Frunzio}},
  \bibinfo{author}{\bibfnamefont{S.~M.} \bibnamefont{Girvin}},
  \bibinfo{author}{\bibfnamefont{M.}~\bibnamefont{Mirrahimi}},
  \bibinfo{author}{\bibfnamefont{M.~H.} \bibnamefont{Devoret}},
  \bibnamefont{and} \bibinfo{author}{\bibfnamefont{R.~J.}
  \bibnamefont{Schoelkopf}}, ``Deterministically encoding quantum information
  using 100-photon {S}chr{\"o}dinger cat states,'' \bibinfo{journal}{Science}
  \textbf{\bibinfo{volume}{342}}, \bibinfo{pages}{607} (\bibinfo{year}{2013}).

\bibitem[{\citenamefont{Sychev et~al.}(2017)\citenamefont{Sychev, Ulanov,
  Pushkina, Richards, Fedorov, and Lvovsky}}]{lvovsky}
\bibinfo{author}{\bibfnamefont{D.~V.} \bibnamefont{Sychev}},
  \bibinfo{author}{\bibfnamefont{A.~E.} \bibnamefont{Ulanov}},
  \bibinfo{author}{\bibfnamefont{A.~A.} \bibnamefont{Pushkina}},
  \bibinfo{author}{\bibfnamefont{M.~W.} \bibnamefont{Richards}},
  \bibinfo{author}{\bibfnamefont{I.~A.} \bibnamefont{Fedorov}},
  \bibnamefont{and} \bibinfo{author}{\bibfnamefont{A.~I.}
  \bibnamefont{Lvovsky}}, ``Enlargement of optical {S}chr{\"o}dinger's cat
  states,'' \bibinfo{journal}{Nat. Photonics} \textbf{\bibinfo{volume}{11}},
  \bibinfo{pages}{379} (\bibinfo{year}{2017}).

\bibitem[{\citenamefont{Lewenstein et~al.}(2021)\citenamefont{Lewenstein,
  Ciappina, Pisanty, Rivera-Dean, Stammer, Lamprou, and Tzallas}}]{lewenstein}
\bibinfo{author}{\bibfnamefont{M.}~\bibnamefont{Lewenstein}},
  \bibinfo{author}{\bibfnamefont{M.~F.} \bibnamefont{Ciappina}},
  \bibinfo{author}{\bibfnamefont{E.}~\bibnamefont{Pisanty}},
  \bibinfo{author}{\bibfnamefont{J.}~\bibnamefont{Rivera-Dean}},
  \bibinfo{author}{\bibfnamefont{P.}~\bibnamefont{Stammer}},
  \bibinfo{author}{\bibfnamefont{T.}~\bibnamefont{Lamprou}}, \bibnamefont{and}
  \bibinfo{author}{\bibfnamefont{P.}~\bibnamefont{Tzallas}}, ``Generation of
  optical `{S}chr{\"o}dinger cats' in intense laser–matter interactions,''
  \bibinfo{journal}{Nat. Phys.} \textbf{\bibinfo{volume}{17}},
  \bibinfo{pages}{1104} (\bibinfo{year}{2021}).

\end{thebibliography}

\end{document}